\documentclass[aps,prd,twocolumn,groupedaddress,nofootinbib,
superscriptaddress,preprintnumbers]{revtex4-1}
\usepackage{epsfig,graphicx}
\usepackage{dcolumn}
\usepackage{bm}
\usepackage{latexsym}
\usepackage{amsfonts}
\usepackage{amssymb}
\usepackage{amsmath}
\usepackage{ascmac}
\usepackage{mathrsfs}

\usepackage{color}

\usepackage[T1]{fontenc} 
\usepackage{tikz-feynhand}
\usepackage{graphicx}
\usepackage{xcolor}
\usepackage{caption,subcaption}
\usepackage{mathrsfs,mathtools}
\usepackage{physics,amssymb}
\usepackage{bm}
\usepackage{braket}
\usepackage{listings}
\usepackage{cases}
\usepackage{comment}
\usepackage{soul}
\usepackage{cancel}
\usepackage{cases}
\usepackage[utf8]{inputenc}
\usepackage{url}
\usepackage[normalem]{ulem}
\usepackage{xspace}
\usepackage{aas_macros}
\usepackage{acronym}
\usepackage{siunitx}

\usepackage{xcolor}
\definecolor{bleudefrance}{rgb}{0.19, 0.55, 0.91}
\definecolor{desyblue}{HTML}{009EE2}
\definecolor{desyorange}{HTML}{FD8800}
\definecolor{dark_red}{rgb}{0.7, 0., 0.}
\definecolor{light_pink}{rgb}{1,0.4,0.4}
\definecolor{lblue}{rgb}{0.384602,0.117763,0.973947}
\usepackage{color}
\input{colordvi.tex}
\allowdisplaybreaks[1]
\usepackage{url}
\usepackage{hyperref}
\hypersetup{
colorlinks=false,
hidelinks}
\newcommand*{\D}{{\rm d}}
\newcommand*{\mpl}{M_{\rm Pl}}

\newcommand*{\mg}[1]{{\color{magenta}#1}}

\definecolor{MONZA}{HTML}{CF000F}
\definecolor{DARKBLUE}{HTML}{00008b}
\definecolor{DARKMAGENTA}{HTML}{8b008b}
\definecolor{DARKCYAN}{HTML}{00cfc0}

\usepackage{soul}

\begin{document}

\title{Bypassing the Lyth Bound with Entangled Gravitons: Primordial Signatures and Late-Time Noise} 

\author{Shingo Akama}
\author{Chunshan Lin}
\affiliation{Faculty of Physics, Astronomy and Applied Computer Science, Jagiellonian University, 30-348 Krakow, Poland}


\begin{abstract}
We demonstrate that quantum entanglement between primordial gravitons in dynamically decoupled gravitational sectors can parametrically enhance the tensor power spectrum during inflation. Unlike standard mechanisms relying on classical dynamics or modified actions, this enhancement originates from the reduced density matrix of the observable sector after tracing over a hidden gravitational reservoir. This framework allows for a sizable tensor-to-scalar ratio $r \gtrsim 0.01$ consistent with sub-Planckian inflaton excursions, providing a purely quantum mechanical evasion of the Lyth bound. The resulting mixed state leaves a distinctive ``quantum birthmark’' in the form of oscillatory features in the power spectrum and  a characteristic violation of the single-field consistency relation, manifesting as a scale-dependent enhancement of the squeezed-limit bispectrum. Furthermore, we forecast that this entanglement may manifest as a late-time stochastic noise enhancement in gravitational wave interferometers, offering a novel experimental window into the quantum nature of spacetime.

\end{abstract}

\maketitle

{\bf Introduction}
Primordial gravitational waves provide a unique probe of the quantum nature of spacetime and the energy scale of inflation. Within the standard single-field slow-roll paradigm, the amplitude of these tensor fluctuations is rigidly tied to the inflaton field excursion. This relationship, known as the Lyth bound~\cite{Lyth:1996im}, implies that a detection of the tensor-to-scalar ratio $r \gtrsim 0.01$ would necessitate super-Planckian field excursions. Such large excursions challenge the validity of the effective field theory description, potentially rendering the theory sensitive to unknown ultraviolet (UV) physics and quantum gravity corrections. Most proposed ways to evade the Lyth bound rely on modifying inflationary dynamics or introducing additional classical sources of tensor modes, such as spectator fields~\cite{Cai:2021yvq}, gauge-field production~\cite{Dimastrogiovanni:2012ew,Adshead:2013qp}, or  modified gravity sectors~\cite{Lin:2015nda,Kunimitsu:2015faa}. Despite their variety, these scenarios share a common structure: the enhancement of tensor fluctuations originates from classical dynamics or classical energy transfer, often accompanied by strong backreaction, enhanced scalar non-Gaussianity, or tight consistency constraints. It is therefore natural to ask whether the Lyth bound can be violated without modifying gravity, altering the inflationary background, or the classical field content.

In this Letter, we demonstrate that such a violation occurs naturally if the observable gravitons are quantum mechanically entangled with a hidden gravitational sector. We consider two dynamically decoupled sectors—our universe and a hidden copy—described by the action $S = S_g + S_f$, where $g_{\mu\nu}$ and $f_{\mu\nu}$ are the metrics of the two sectors. A compelling motivation for this setup can be the quantum creation of a multiverse~\cite{Sato:1981gv,Vilenkin:1983xq,Linde:1986fc,Linde:1986fd,Bousso:2000xa,Susskind:2003kw}. Within the framework of the Hartle-Hawking no-boundary proposal \cite{Hartle:1983ai} or the tunneling-from-nothing scenarios \cite{Vilenkin:1982de,Vilenkin:1983xq}, universes may be nucleated in entangled pairs bridged by a Euclidean wormhole \cite{Bouhmadi-Lopez:2017sgq,Chen:2025kwx}. Such geometric connectivity is supported by the study of Euclidean wormholes (see, e.g., Ref. \cite{Hebecker:2018ofv}), where entangled pairs of universes are realized through a double Euclidean instanton configuration Refs.~\cite{Robles-Perez:2011yds,Garay:2013pba,Robles-Perez:2013kva}.
Following the ER  = EPR  conjecture~\cite{Maldacena:2013xja}, this geometric connectivity manifests as quantum entanglement between the degrees of freedom in the two now-disconnected spacetimes. 
Although the two universes evolve independently at the level of the low-energy effective action, their shared quantum origin prepares the primordial gravitons in an entangled state. Tracing over the unobserved hidden sector modifies the reduced density matrix of the observable gravitons, driving them into a mixed quantum state. This is fundamentally distinct from the pure excited states (such as non-Bunch-Davies states) typically considered in single-universe models.\footnote{Similar to the scalar field case studied in Refs.\cite{Kanno:2015lja,Kanno:2015ewa,Rostami:2017ktl,Bolis:2018jmo}, the global entangled state can be formally constructed via a two-mode Bogoliubov transformation acting on the product vacuum. We emphasize, however, that the resulting density matrix for the observable sector is mixed and cannot be purified by a local Bogoliubov transformation restricted to a single observable metric.}

While our setup is reminiscent of bimetric gravity (bigravity), it is conceptually and technically distinct. Standard bigravity, although phenomenologically rich, faces formidable theoretical challenges. Constructing a consistent interaction between two metrics requires a highly specific, non-linear potential to avoid the Boulware-Deser ghost~\cite{deRham:2010kj}. Even in ``ghost-free’' formulations, cosmological solutions are frequently plagued by gradient instabilities at early times~\cite{Koennig:2014ods} or require fine-tuning to satisfy the Higuchi bound while maintaining Vainshtein screening~\cite{Fasiello:2012rw}. Furthermore, the complexity of the interaction potential often obscures the physical origin of the signal.  In contrast to the bigravity theory whose two metrics couple in the classical action,  we couple them in the quantum state, and therefore, fortunately, all these issues are absent. 

The resulting phenomenology is qualitatively distinct from that of excited or non-Bunch-Davies initial states within a single sector: the enhancement acts as a fundamental mixed-state contribution that cannot be removed by a local Bogoliubov transformation of the observable modes and is accompanied by characteristic oscillatory features in the tensor power spectrum, together with distinctive modifications of tensor non-Gaussianity and single-field consistency relations. Crucially, if the large tensor-to-scalar ratio accompanied with oscillatory features and correlations are confirmed and varied across future experiments, they would constitute direct evidence that primordial tensor modes originate from genuinely quantum degrees of freedom, providing empirical support for the quantization of the gravitational field itself. In this way, a large tensor-to-scalar ratio can coexist with an arbitrarily small inflaton excursion, yielding a parametric violation of the Lyth bound, while elevating primordial gravitational waves to probes of quantum entanglement and of the quantum nature of gravity, with potentially transformative impact on cosmology and fundamental physics.

{\bf The primordial Entangled Graviton State }
We consider a bipartite system of dynamically decoupled gravitational sectors, $g_{\mu\nu}$ and $f_{\mu\nu}$, governed by the total action $S = S_g + S_f$. In this framework, $g_{\mu\nu}$ represents the physical metric coupled to the observable matter sector and undergoes inflationary phase, while $f_{\mu\nu}$ describes a hidden gravitational sector. For the purposes of this study, we assume that the hidden gravitational sector resides in a pure de Sitter background with a bare cosmological constant $\Lambda_{\text{hidden}}$, and acts as a quantum reservoir for the observable tensor modes. Let us define the tensor perturbations as
$g_{ij}=a_g^2(\delta_{ij}+h_{ij})$
and $f_{ij} =a_f^2(\delta_{ij}+\gamma_{ij}).$ While the sectors do not interact during the expansion, they share a quantum history that prepares them in an entangled state at an initial conformal time $\eta_0$. The tensor perturbations are decomposed into Fourier modes $h_{\mathbf{k}, s}$ and $\gamma_{\mathbf{k}, s}$, where $s = (+, \times)$ denotes polarization. At quadratic order, the graviton action is given by
\begin{align}\label{action2}
S^{(2)}_{\rm grav}&=\frac{\mpl^2}{8}\int\D\eta\D^3 xa_g^2(\eta)\left[(h_{ij}')^2-(\partial_k h_{ij})^2\right]\nonumber\\
&~ +\frac{M_f^2}{8}\int\D\eta\D^3 xa_f^2(\eta)\left[(\gamma_{ij}')^2-(\partial_k \gamma_{ij})^2\right]. 
\end{align}

The initial state of the system is described by a Gaussian wave functional $\Psi[h, \gamma]$ that,  in the absence of interactions in the quadratic action, can be factorized into the product of wave functional of each ${\bf k}$-mode, namely, $\Psi=\prod_k\psi(\textbf{k})$ where
\begin{eqnarray}
&&\psi({\bf k})=N_k\exp\biggl\{-\frac{1}{2}\sum_{s,s',\sigma,\sigma'}\left[A^{(ss')}_k h^{(s)}_{{\bf k}}h^{(s')}_{-{\bf k}}\right.\nonumber\\
&&\left.+B^{(\sigma\sigma')}_k \gamma^{(\sigma)}_{{\bf k}}\gamma^{(\sigma')}_{-{\bf k}}+C_k^{(s\sigma)}\left(h^{(s)}_{\bf k}\gamma^{(\sigma)}_{-{\bf k}}+h^{(s)}_{{\bf k}}\gamma^{(\sigma)}_{-{\bf k}}\right)\right]\biggr\},~~~~
\end{eqnarray}
where $C_{\bf k}^{(s\sigma)}$ characterizes the entanglement between the two graviton sectors, while  the normalization condition fixes $N_k$.
The wave functional obeys the Schr$\text{\"{o}}$dinger equation,
\begin{align}\label{sdgeq}
i\partial_\eta\Psi=\hat H\Psi,
\end{align}
where the Hamiltonian $\hat H$ is obtained by performing the Legendre transformation of the action~(\ref{action2}) and promoting the conjugate momenta of $h_\textbf{k}$ and $\gamma_\textbf{k}$ to operators. Eq. (\ref{sdgeq}) represents a set of coupled differential equations for the Gaussian kernels $A^{(ss')}_k$, $B^{(\sigma\sigma')}_k$ and $C_k^{(s\sigma)}$. In what follows, we adopt a simplifying assumption that the entanglement between the two polarization states $h_k^+$ and $h_k^\times$ (as well as $\gamma_k^+$ and $\gamma_k^\times$) can be neglected, as such mixing arises only through nonlinear graviton interactions and is therefore subleading relative to the leading-order effects considered here. We also restrict attention to the diagonal components of the entanglement kernel $C_k^{(s\sigma)}$, neglecting off-diagonal contributions. This allows us to isolate the essential physical consequences of graviton entanglement while deferring a more general and exhaustive treatment to future work.  In passing, we shall mention that we would expect similar results in more general cases such that $A^{(ss')}_k, B^{(ss')}_k, C^{(ss')}_k \neq 0$ for $s \neq s'$.

By killing the off-diagonal parts of kernels, we obtain
\begin{align}
i[A^{(ss)}_k]'&=\frac{1}{z_g^2}[A^{(ss)}_k]^2+\frac{1}{z_f^2}[C^{(ss)}_k]^2-z_g^2k^2,\\
i[B^{(ss)}_k]'&=\frac{1}{z_f^2}[B^{(ss)}_k]^2+\frac{1}{z_g^2}[C^{(ss)}_k]^2-z_f^2k^2,\\
i[C^{(ss)}_k]'&=\biggl(\frac{1}{z_g^2}A^{(ss)}_k+\frac{1}{z_f^2}B^{(ss)}_k\biggr)C^{(ss)}_k.
\end{align}
where $z_g\equiv a_g\mpl/2$ and $z_f\equiv a_fM_f/2$. These Riccati-type equations for $A_k$ and $B_k$ can be linearized by introducing auxiliary mode functions $g_k^{(s)}$ and $f_k^{(s)}$, where $iA^{(ss)}_k\equiv z_g^2\ln{\left(g_k^{(s)}/z_g\right)}'$ and $iB^{(ss)}_k\equiv z_f^2\ln{\left(f_k^{(s)}/z_f\right)}'$. This mapping transforms the complex kernel dynamics into a pair of coupled, second-order differential equations 
\begin{align}
[g^{(s)}_k]''+\biggl(k^2-\frac{a_g''}{a_g}\biggr)g_k^{(s)}&=\frac{[\lambda_{k}^{(s)}]^{2}}{g_{k}^{(s)}[f_{k}^{(s)}]^{2}},\\
[f^{(s)}_k]''+\biggl(k^2-\frac{a_f''}{a_f}\biggr)f_k^{(s)}&=\frac{[\lambda_{k}^{(s)}]^{2}}{[g_{k}^{(s)}]^{2}f_{k}^{(s)}} .
\end{align}
where $\lambda_k^{(s)}$ is a complex parameter quantifying the initial UV entanglement between the two sectors as $C_k^{(ss)}=\lambda_k^{(s)}z_gz_f/(g_k^{(s)}f_k^{(s)})$. 
Notably, the presence of entanglement introduces a non-linear ``source’' term even in the absence of a classical coupling in the action.

In standard inflation, the two-point function represents the variance of zero-point fluctuations in a pure Bunch-Davies vacuum. However, in our case it undergoes a profound shift in meaning: it measures the stochastic noise originating from a hidden sector! The observable universe is thus in a mixed state upon tracing over the hidden sector $\gamma_k$,
\begin{equation}
\langle h^{(s)}_{\bf k}h^{(s')}_{\bf k'}\rangle=\frac{1}{\mathcal{N}}\int\left[\prod_{\sigma,s''}\mathcal{D}^2h^{(s'')}_{\bf q}\mathcal{D}^2\gamma^{(\sigma)}_{\bf q}\right]h^{(s)}_{\bf k}h^{(s')}_{\bf k'}|\Psi|^2,
\end{equation}
where
$\mathcal{N}:=\int\left[\prod_{\sigma,s}\mathcal{D}^2h^{(s)}_{\bf q}\mathcal{D}^2\gamma^{(\sigma)}_{\bf q}\right]|\Psi|^2$.
 Physically, the tracing operation introduces a ``noise’' term, a stochastic bath of fluctuations provided by the hidden sector in which our observable gravitons being immersed. This ``bath’' increases the variance (the amplitude) of the modes without requiring a higher energy density from the inflaton field itself. The resulting two-point function is
\begin{eqnarray}\label{tensorPS}
\langle h^{(s)}_{\bf k}h^{(s)}_{-\bf k}\rangle&=\frac{|g^{(s)}_k|^2}{z_g^2}\{1-4[\lambda^{(s)}_k]^2\cos^2(\theta_{gs}+\theta_{f s})\}^{-1},~~
\end{eqnarray}
where $g^{(s)}_k\equiv|g^{(s)}_k|e^{i\theta_{gs}}$ and $f^{(s)}_k\equiv|f^{(s)}_k|e^{i\theta_{f s}}$ are the phases of the mode functions in the two sectors. The term $\cos^2(\theta_{gs}+\theta_{fs})$ represents a ``quantum birthmark’': a characteristic oscillatory interference pattern in the power spectrum that serves as a direct signature of the pre-inflationary entanglement between the two universes. We hereafter assume that $\lambda_k^{(s)}$  is a real quantity for simplicity, and it satisfies the condition $-1/2<\lambda^{(s)}_k<1/2$ so that the wave functional can be normalized.

\begin{figure} [htb]
     \begin{tabular}{cc}
        \begin{minipage}{0.49\hsize}
            \centering
            \includegraphics[width=4.6cm]{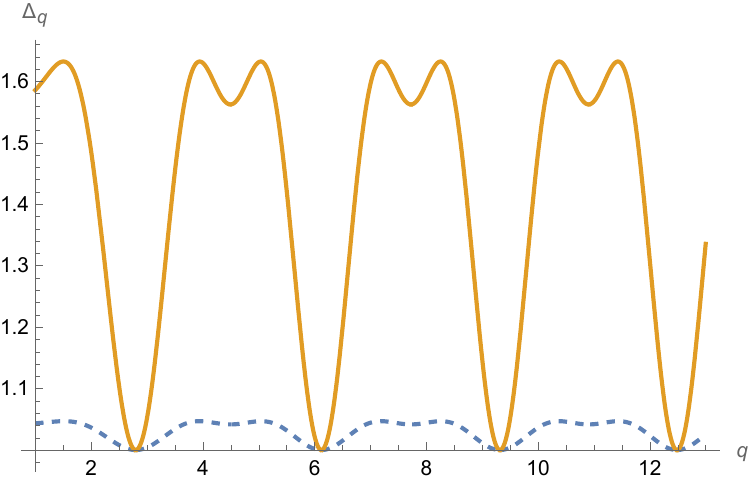}
        \end{minipage} &
        \begin{minipage}{0.49\hsize}
            \centering
            \includegraphics[width=4.6cm]{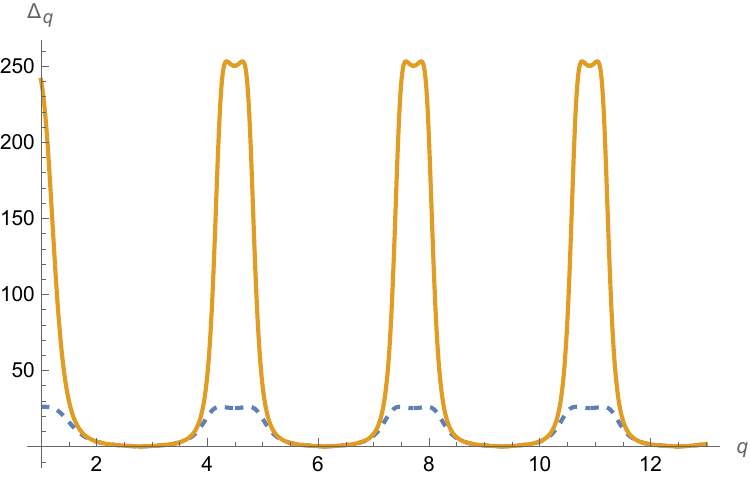}
        \end{minipage} 
    \end{tabular}
\caption{%
\raggedright\emph{Left}: The plot of $\Delta_q$ for $\lambda^{(s)}_k=0.1$ (dashed line) and $\lambda^{(s)}_k=0.3$ (thick line)
. \emph{Right}: The plot of $\Delta_q$ for $\lambda^{(s)}_k=0.49$ (dashed line) and $\lambda^{(s)}_k=0.499$ (thick line). 
Slow-roll corrections are neglected for all cases.}\label{Fig: Delta}
\end{figure}

A key result of this Letter is the parametric enhancement of the tensor power spectrum in the limit of maximal entanglement  $|\lambda_{k}^{(s)}|\rightarrow1/2$, where the spectrum formally diverges.  This amplification allows for a significant enhancement of the tensor-to-scalar ratio $r$ even at a low inflationary energy scale $H$. To quantify this, we define the amplification factor $\Delta_q$ relative to the standard Bunch-Davies spectrum,
\begin{align}
\Delta_q:=\frac{\langle h^{(s)}_{\bf k}h^{(s)}_{-\bf k}\rangle}{\langle h^{(s)}_{\bf k}h^{(s)}_{-\bf k}\rangle_{BD}}\biggr|_{\eta=\eta_f}
\end{align}
where $\langle h^{(s)}_{\bf k}h^{(s)}_{-\bf k}\rangle_{BD}$ is the standard inflationary tensor power spectrum evaluated in a Bunch-Davies vacuum, and the ratio is computed at the end of the inflation when $\eta=\eta_f$. We show $\Delta_q$ as a function of
$q:=-k\eta_0$ for low frequency modes, which are initially around horizon crossing (i.e., $q=\mathcal{O}(1)$), for the purpose of visualizing deviations from the Bunch-Davies case.
A similar plot is obtained for high frequency modes, which are initially on subhorizon scales (i.e., $q\gg1$).
Numerical evaluation (Fig.~\ref{Fig: Delta}) shows that for $\lambda_k^{(s)} \to 0.499$, the power spectrum is enhanced by several orders of magnitude. 
Consequently, the field excursion required for an observable $r$ is rescaled as:
\begin{align}
\Delta\phi\sim\sqrt{2\epsilon}N\mpl\sim 2\{1-4[\lambda^{(s)}_k]^2\}^{1/2}\mpl\biggl(\frac{r}{0.01}\biggr)^{1/2}\biggl(\frac{N}{60}\biggr),
\end{align}
where we neglected the time variation of $\epsilon$ and used
$N=\int \frac{H}{\dot \phi}\D\phi\simeq  \frac{1}{\sqrt{2\epsilon}}\frac{\Delta\phi}{\mpl}$,
with $N$ being the e-folding number. The scalar power spectrum was assumed to be identical to that in the Bunch–Davies case, since scalar–tensor mixing is typically a subleading effect. By taking $|\lambda_{k}^{(s)}|\rightarrow 1/2$, the field excursion can be suppressed to arbitrarily sub-Planckian values, $\Delta \phi \ll M_{Pl}$, effectively evading the Lyth bound through purely quantum mechanical means. This demonstrates that large-amplitude primordial gravitational waves do not necessarily imply large-field inflation, but may instead point toward a rich, entangled pre-inflationary history of the multiverse.


Apart from the amplitude enhancement, another striking feature is the high-frequency oscillatory behavior (which was also found in previous works~\cite{Albrecht:2014aga,Bolis:2016vas,Collins:2016ahj,Baunach:2023okd}).
These oscillations originate from the $\cos^2(\theta_{gs} + \theta_{fs})$ term in the two-point function. Physically, they represent the constructive and destructive interference between the observable $g$-modes and the hidden $f$-modes at the moment of their shared creation or horizon crossing. Unlike standard non-Bunch-Davies states, which often produce smooth shifts in power, this entanglement-induced interference creates a distinct ``fringe pattern’' across the spectrum. The frequency and phase of these oscillations are determined by the initial conformal time $\eta_0$ and the relative evolution of the two scale factors, $a_g$ and $a_f$. Consequently, these features serve as a robust observational signature: a detection of such periodic modulations in the primordial gravitational wave spectrum would not only point toward a non-vacuum initial state but would specifically identify a history of quantum connectivity between our universe and a hidden gravitational sector.


The quantum signature of this entangled state extends beyond the power spectrum to higher-order correlations.
In standard single-field inflation with a Bunch-Davies vacuum, the primordial tensor bispectrum in the squeezed limit is rigidly constrained by Maldacena’s consistency relation~\cite{Maldacena:2002vr}. However, our framework deviates profoundly from this benchmark. Drawing from the analogous scalar analysis in Ref. ~\cite{Bolis:2019fmq}, we find that the tensor bispectrum scales as $\{1-4[\lambda^{(s)}_k]^2\}^{-3}$, leading to an overall amplification of the non-linearity parameter $f_{\rm NL}$ by a factor of $\{1-4[\lambda^{(s)}_k]^2\}^{-1}$. Physically, the long-wavelength modes of the hidden sector act as a modulated environment for the short-wavelength observable modes, inducing a scale-dependent enhancement in the squeezed configuration ($k_L \ll k_S$) such that $f_{\rm NL} \propto k_S/k_L$. This explicit violation of the Maldacena’s consistency relation  provides a robust discriminatory test:  while the
power spectrum enhancement facilitates the evasion of the Lyth bound, the characteristic enhancement and scale-dependence of
the bispectrum serve as a “smoking gun” to distinguish this quantum entanglement mechanism from classical modifications of gravity or standard excited initial
states. A detailed derivation of these results will be presented in our forthcoming work~\cite{Akama:prep}.

{\bf Late-time Implications}
If the quantum entanglement between the two sectors persists into the late-time universe, it offers a unique phenomenological profile: it preserves classical macroscopic gravity while parametrically enhancing fundamental quantum fluctuations. 

A key distinction of our model compared to classical bimetric theories is that the modification resides in the quantum state, not the action. The static Newtonian potential between two masses is determined by the tree-level exchange amplitude of a virtual graviton, governed by the Feynman propagator $D_{\mu \nu \rho \sigma}(k)$. Since the action $S = S_g + S_f$ contains no cross-terms and remains the standard Einstein-Hilbert action for the observable sector, the propagator remains $D(k) \propto 1/k^2$. Consequently, the static potential remains strictly Newtonian:$V(r) = -G M_1 M_2/r$. This ensures that the model automatically satisfies Solar System and laboratory tests of gravity without requiring screening mechanisms. The entanglement does not manifest as a ``fifth force’', but rather as a modification of the underlying quantum noise.

While the classical response is unchanged, the mixed-state nature of the gravitons leads to a significant modification of the stochastic noise kernel. Following the formalism of Parikh, Wilczek, and Zahariade~\cite{Parikh:2020kfh,Parikh:2020fhy}, the separation $\xi(t)$ of two falling masses (such as the mirrors of an interferometer) is subject to a stochastic Langevin-like correction: $\ddot{\xi}(t) \approx \frac{1}{2} \xi_0 \left[ \ddot{h}_{cl}(t) + \ddot{N}(t) \right]$, where $N(t)$ represents the fundamental noise induced by the quantum fluctuations of the metric. The statistics of this noise are determined by the symmetrized two-point correlator of the graviton field. Vacuum or coherent states give Planck-suppressed noise, while squeezed states give exponential enhancement.

In our framework, however, the reduced density matrix is mixed even if the global state is pure. The noise kernel acquires an extra contribution from entanglement. Tracing over the hidden sector $\gamma_k$ generates a non-Markovian stochastic bath.
 The stochastic force on observables is enhanced without requiring large occupation numbers or strong squeezing. We forecast that the noise power spectral density $S_h(\omega)$ is parametrically enhanced by the entanglement strength, following the scaling of the two-point function:  $S_h(\omega) \sim S_{vac}(\omega) /\left[1 - \mathcal{O}\left(\lambda^2\right)\right]$, where $\lambda$ is the late-time limit of the entanglement parameter (which may differ from the primordial one). While the approach to the maximal limit $\left[1 - \mathcal{O}\left(\lambda^2\right)\right] \to 0$ may be regulated by backreaction or decoherence effects, the enhancement remains significant in the large-entanglement regime. Crucially, this noise carries a specific ``birthmark’': the oscillatory $\cos^2(\dots)$ features predicted in our primordial analysis. These oscillations, tied to the initial entanglement scale, provide a distinctive signature that cannot be replicated by instrumental or astrophysical noise sources. Detection of such a signal in next-generation detectors like LISA~\cite{LISA:2017pwj} or the Einstein Telescope~\cite{Punturo:2010zz} could provide the first direct experimental evidence for quantum entanglement in the gravitational field.

 {\bf Conclusion and Discussion }
In this Letter, we have demonstrated that quantum entanglement between dynamically decoupled gravitational sectors provides a novel mechanism to enhance primordial tensor fluctuations without modifying the inflationary background or the classical gravitational action. By tracing over a hidden gravitational sector—motivated by the quantum nucleation of entangled universe pairs—the observable gravitons are driven into a mixed state characterized by a modified reduced density matrix. This framework allows the tensor-to-scalar ratio $r$ to reach observable levels even for sub-Planckian field excursions, thereby providing a purely quantum mechanical evasion of the Lyth bound.

Our results reveal that the entangled state leaves a distinctive ``quantum birthmark’' in the form of oscillatory features in the tensor power spectrum. Furthermore, we forecast a specific violation of the single-field consistency relation, where the bispectrum in the squeezed limit is parametrically enhanced by the effect of entanglement.
These features are not merely technical artifacts; they represent a fundamental departure from the vacuum fluctuations of standard inflation and cannot be replicated by classical modifications of gravity or standard non-Bunch-Davies states within a single sector.


Extending this analysis to the late-time universe, we find that while the static Newtonian potential remains unchanged—ensuring consistency with local gravitational tests—the fundamental quantum noise in the metric is significantly amplified. This stochastic ``jitter’' carries the same oscillatory signature as the primordial spectrum. Consequently, next-generation gravitational wave interferometers such as LISA and the Einstein Telescope may serve as direct probes of quantum entanglement in gravity. A detection of the predicted spectral oscillations would not only confirm the existence of a hidden sector but would also constitute empirical evidence for the quantization of the gravitational field itself. By elevating primordial gravitational waves to probes of the quantum connectivity with the hidden, this work opens a new window into the intersection of quantum information and fundamental cosmology.

{\bf Acknowledgment}
We would like to thank G. Zahariade and O. Shetye for discussions. 
This work was supported by the grant No.~UMO-2021/42/E/ST9/00260 from the National Science Centre, Poland.

\bibliography{letter}

\end{document}